# Automatic Phone Slip Detection System

Karthik R, Preetam Satapath, Srivatsa Patnaik, Saurabh Priyadarshi, Rajesh Kumar M, *Senior Member, IEEE*
Department of Electronics and Communication Engineering VIT University Vellore, India 632014
tkgravikarthik@gmail.com, mrajeshkumar@vit.ac.in

*Abstract*— Mobile phones are becoming increasingly advanced and the latest ones are equipped with many diverse and powerful sensors. These sensors can be used to study different position and orientation of the phone which can help smartphone manufacture to track about their customers handling from the recorded log. The inbuilt sensors such as the accelerometer and gyroscope present in our phones are used to obtain data for acceleration and orientation of the phone in the three axes for different phone vulnerable position. From the data obtained appropriate features are extracted using various feature extraction techniques. The extracted features are then given to classifier such as neural network to classify them and decide whether the phone is in a vulnerable position to fall or it is in a safe position .In this paper we mainly concentrated on various case of handling the smartphone and classified by training the neural network.

## 1. Introduction

Human activity recognition system using devices like cameras or microphones have become an active field of research. It has the potential to be applied in different applications such as ambient assisted living. So human activity recognition system has become a part of our daily lives. Smartphones incorporates many diverse and powerful sensors, which can be used for human activity recognition. These sensors include GPS sensors, audio sensors (microphone), vision sensors (cameras), temperature sensors, acceleration sensors (accelerometers), light sensors and direction sensors (magnetic compasses). The data from these sensors can be transferred through wireless communication such as Wi-Fi, 4G and Bluetooth [1]. We have seen that accelerometers and gyroscopes have the most applications as they are the most accurate ones [2].

Studies have shown that activity recognition system using mobile phones are the most extensively used topic in the research domain. Motion-based and location-based activity recognition using in-built sensor and wireless transceivers are the dominating type of activity recognition on mobile phones [3]. Under motion-based activity recognition systems, 3-axis accelerometers are the mostly used sensors available on phones. Most of the studies focus on detecting the locomotion activities, such as standing, walking [4]. Study on various phone positions and orientations and how these positions change different parameters of in-built gyroscope and accelerometer has been limited [5] [6].When a phone is kept at a particular orientation or position there many parameters associated to it. The location of the phone decides a lot about its future [7].

Good results have been obtained in [8] that uses Ameva discretization algorithm and a new Ameva-based classification system to classify physical activity recognition on Smartphones. On comparing the accuracy of human activity recognition, it was found that using only a basic accelerometer gave an accuracy of 77.34%. However, this ratio increased to 85% when basic features are combined with angular features calculated from the orientation of the phone [9]. Human activity recognition using accelerometer was done for some common positions and accuracy was around 91% [10]. Hence an accelerometer alone cannot give very accurate results. But activity recognition significantly increases the efficiency. In contradiction, [11] suggests that this technique still needs a lot of research before it can be used for the general masses. Using few pre-processing techniques, efficiency can be increased too but with many limitations [12].

In this paper, we have focused on different phone positions which are considered to be risky and harmful. Based on these risky positions various parameters such as the roll, pitch and azimuth changes. Whether the phone is at a slipping point or kept on a table or kept on a book. All these factors would decide if the phone will be safe after a certain movement or jerk is applied. And if the jerk moves the phone by a certain distance, will the phone be still safe or there would be a wide change in orientation of the phone which may result in the fall of it. To know all these things beforehand, we have come up with an idea which will tell the user by the change in orientation of the phone that whether the phone is safe or it is in a risky position. The most basic sensors that can be used for these cases are accelerometer and gyroscope. So here we selected some cases which includes normal touch, accident keep, complete slip, slip till tipping point, flip and fall. For all these six positions, 20 samples each are taken. The acceleration and orientation values for each sample are stored.

The data obtained can be plotted which can be filtered and the appropriate features can be extracted [13]. Based on the features extracted, classification algorithm can be implemented using machine learning so that the system can automatically classify different positions. From [14], we got to know that people do consider many aspects before placing a phone somewhere. So based on those aspects, we selected the various positions of the phone.

In section II, the methodology of how data was collected for various samples of different phone slip cases and also procedure to generate the required database is discussed. In Section III, the procedure followed to extract features from the created database is discussed. In section IV, the method used to create a database from the extracted feature is mentioned. In section V, various machine learning classification algorithms is discussed and also how these algorithms are used to classify various phone slip cases is discussed with the observations obtained after implementing the various classification algorithms is discussed and tabulated. In section VI, the final conclusion is drawn depending upon the results obtained.

## 2. Creation of Database

In our study, we considered six phone slipping cases: normal touch keep case, accidental keep case, complete slipping case, slip till tipping point case, flipping case and falling case. The following phone slipping case was chosen as these are the most common ways by which a phone is vulnerable to fall or slip. The first case which is the normal touch keep case is the reference case where the phone is just kept on table by the user and the observations is recorded while performing this act. The second case is the accidental keep where the phone is thrown on a table or chair in a violent way. The next case taken into consideration is the complete slip case where the phone is made to slip completely down a slope. The next case is slip till tipping point case where the phone is placed on a slope and the reading are recorded until the point the phone starts slipping. In the flipping case, the phone was flipped and thrown from one point to another. The last case is the falling case where the phone was subjected to a controlled fall from different heights.

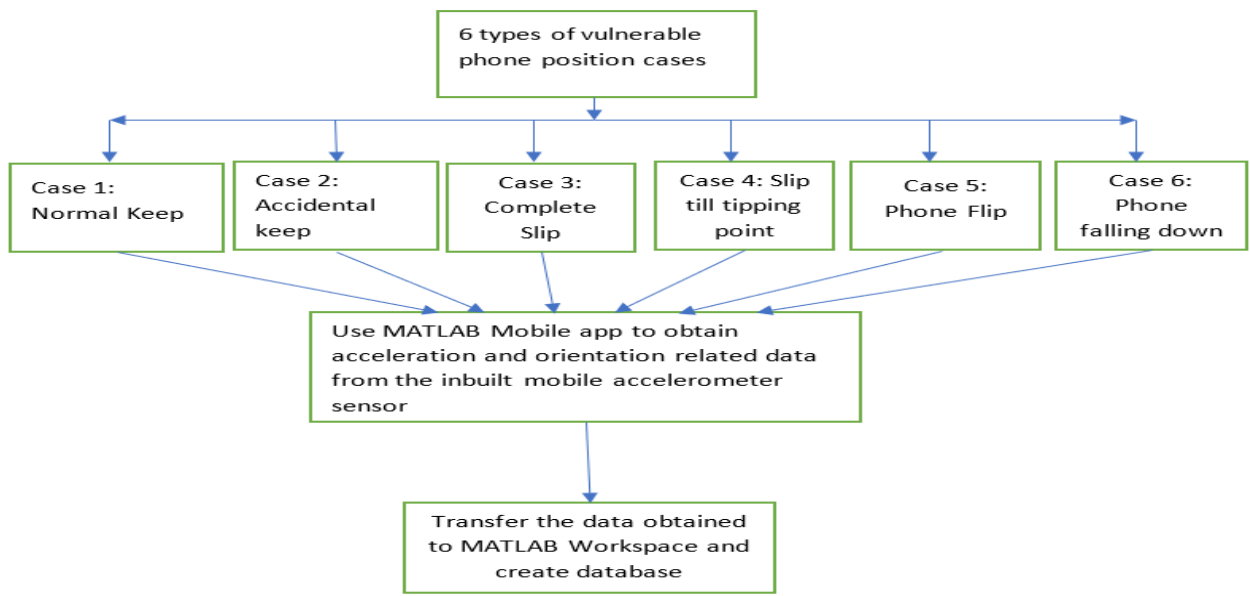

Fig .1 Phone position cases

In order to create the required database, 20 samples were taken for each of the phone slipping case. As there were six cases, totally 120 samples. To collect the samples for each case, MATLAB mobile APP was used. For each the samples of the different cases, data was collected for the acceleration on x, y and z axes and also on the orientation- azimuth, pitch and roll [15]. The data for the above were acquired using the built-in sensors-accelerometer and gyroscope in the device. The built-in accelerometer present in the phone is the device that measures phone acceleration which is the rate of change of velocity of the phone. These accelerometers can measure acceleration in one, two or three axes namely the x, y and z axes. The orientation of the phone is also measured by the accelerometer. The phone orientation includes three parameters- azimuth, pitch and roll. The azimuth is the degrees of rotation about the z axis. This is the angle between the device current compass direction and magnetic north. If the top edge of the phone faces magnetic north, the azimuth is 0 degrees. The second parameter pitch is the degrees of rotation about the x axis. This is the angle between a plane parallel to the device screen and a plane parallel to the ground. The final orientation parameter is the roll which is the degree of rotation about the y axis. This is the angle between a plane perpendicular to the device's screen and a plane perpendicular to the ground. The raw data has been collected using MATLAB 2017a and MATLAB mobile APP  The following graphs shows us the various graph plots of various phone slip cases for the acceleration and orientation related parameters

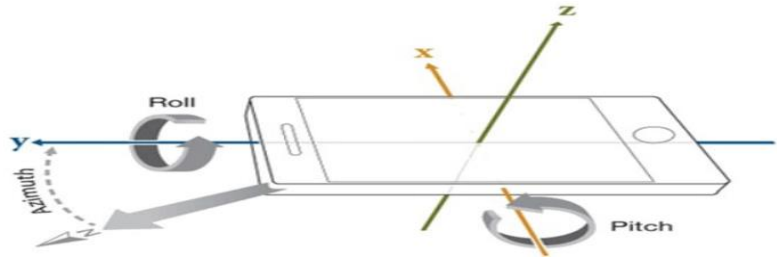

Fig.2 Angular projection of phone [16]

The gyroscope dates are used to determing the rotational motion of an object by measuring angular rate which includes Roll ,pitch and azimuth ,values varies from positive to negative .The gyroscope reading are neither afffetced nor influnced by gravityThe reading from accelerometer are of measurement of linear motin and gravity concurrently .It has influnce of gravity and sensitive to movement which results in acceleration forces

Fig.3 and fig.4 , Fig.5 and fig.6 , Fig.7 and fig.8, Fig.9 and fig.10 , Fig.11 , Fig.12, Fig.13 and Fig.14  shows the sample accelerometer and gyroscope graph of Normal touch ,Accident keep, Complete slip, Slip till tipping point ,Flip and Fall respectively .

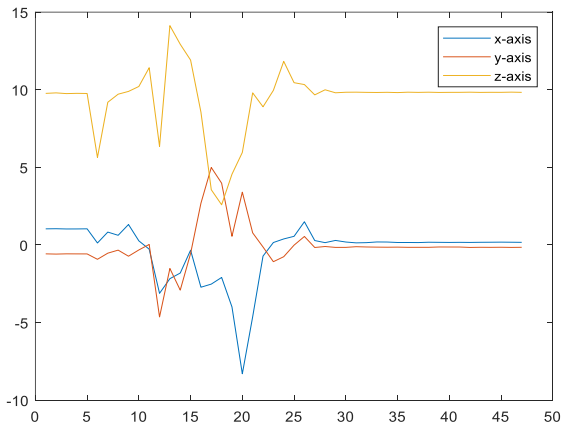

Fig.3 Normal Touch keep-accelerometer reading

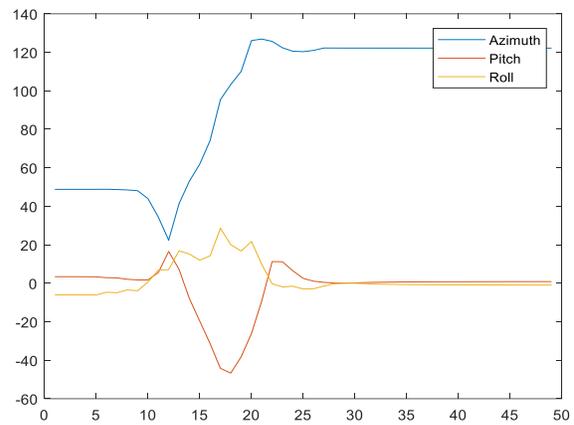

Fig.4 Normal Touch keep-Gyroscope reading

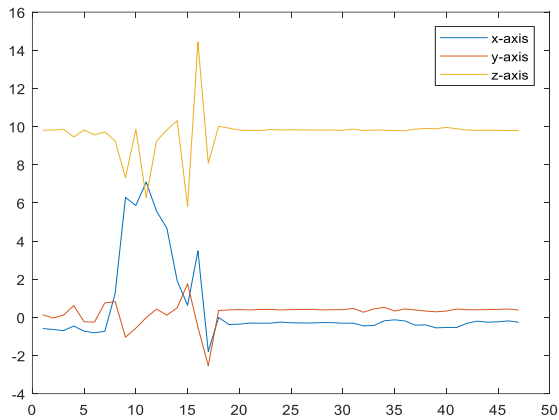

Fig.5 Accident Keep-accelerometer reading

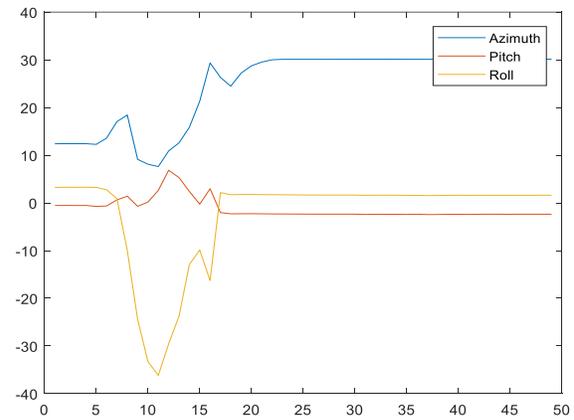

Fig 6 Accident Keep-Gyroscope reading

Fig3 and Fig4 show the accelerometer readings vs time in the three different axes and the gyroscope readings vs time for the first case- Normal touch keep. Fig.5 and Fig.6 show the accelerometer readings vs time in the three different axes and the gyroscope readings vs time for the second case- Accidental keep respectively.

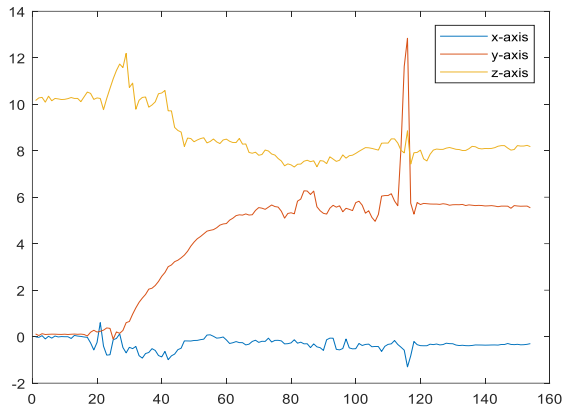
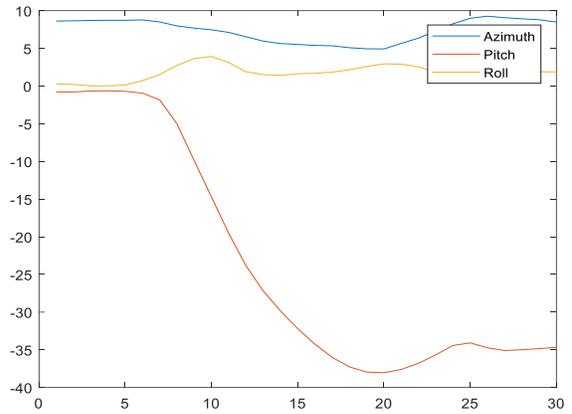

Fig.7 complete Slip-accelerometer reading                Fig.8 Complete Slip-Gyroscope readings

Fig.7 and Fig.8 show the accelerometer readings vs time in the three different axes and the gyroscope readings vs time for the third case- Complete Slip respectively.

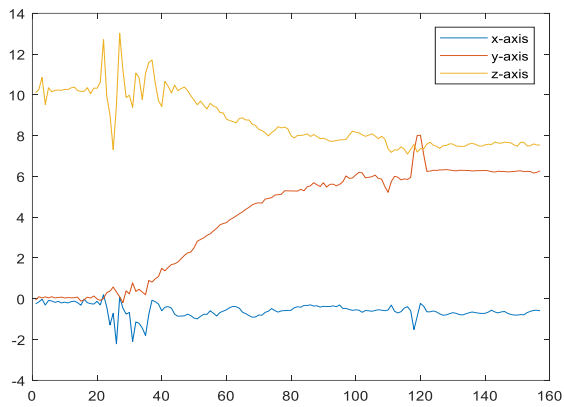
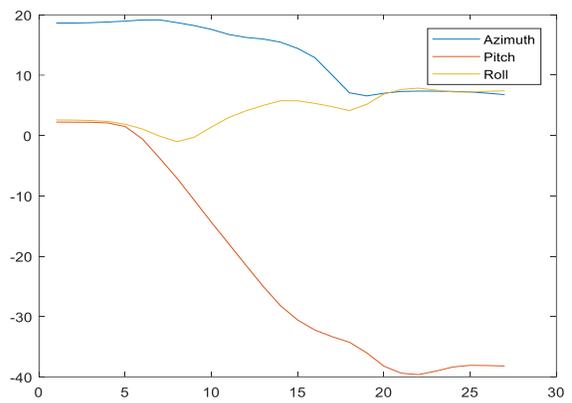

Fig.9 Slip till tipping point-accelerometer reading            Fig.10 Slip till tipping point-Gyroscope reading

Fig.9 and Fig.10 show the accelerometer readings vs time in the three different axes and the gyroscope readings vs time for the fourth case- Slip till Tipping Point respectively

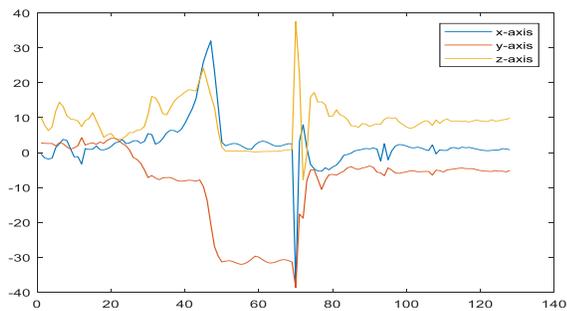
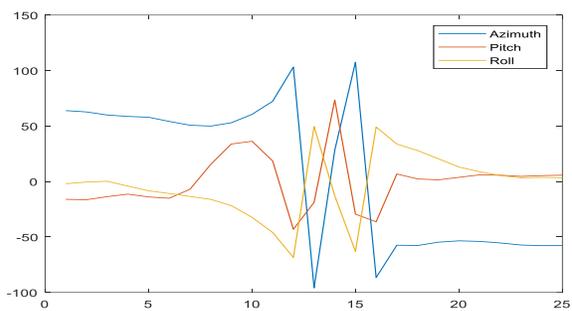

Fig.11 Flip-accelerometer reading                Fig.12 Flip-Gyroscope reading

Fig.11 and Fig.12 show the accelerometer readings vs time in the three different axes and the gyroscope readings vs time for the fifth case- Flip respectively.

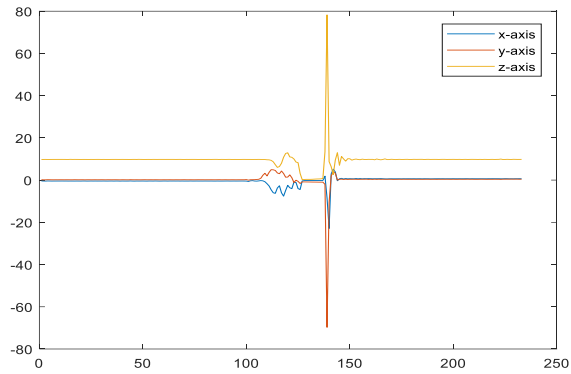
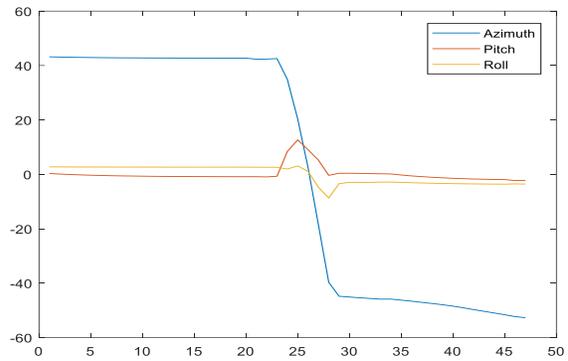

Fig.13 Fall-accelerometer reading                    Fig.14 Fall-Gyroscope reading

Fig.13 and Fig.14 show the accelerometer readings vs time in the three different axes and the gyroscope readings vs time for the sixth case- fall respectively.

The data obtained from the phone was made into a database for further use. This database was subjected to various feature extraction procedures. As a result of which, necessary and important features were obtained. These features were used in various classification neural network algorithms in order to classify them into their categories.

### 3. FEATURE EXTRACTION

One of the main objective of this paper is the extraction of various features from the database of different phone slip cases created. The database created consisted of 10 different samples for each of the 6 different phone slip cases. From these samples, the required features and parameters are extracted for further classification purposes. As the database created is a rather vast. Therefore, it needs to be transformed into a set of reduced set of features. As a result of which these extracted features are expected to contain the relevant information from the given database. Further classification algorithms can be applied on these features instead of the complete initial data.

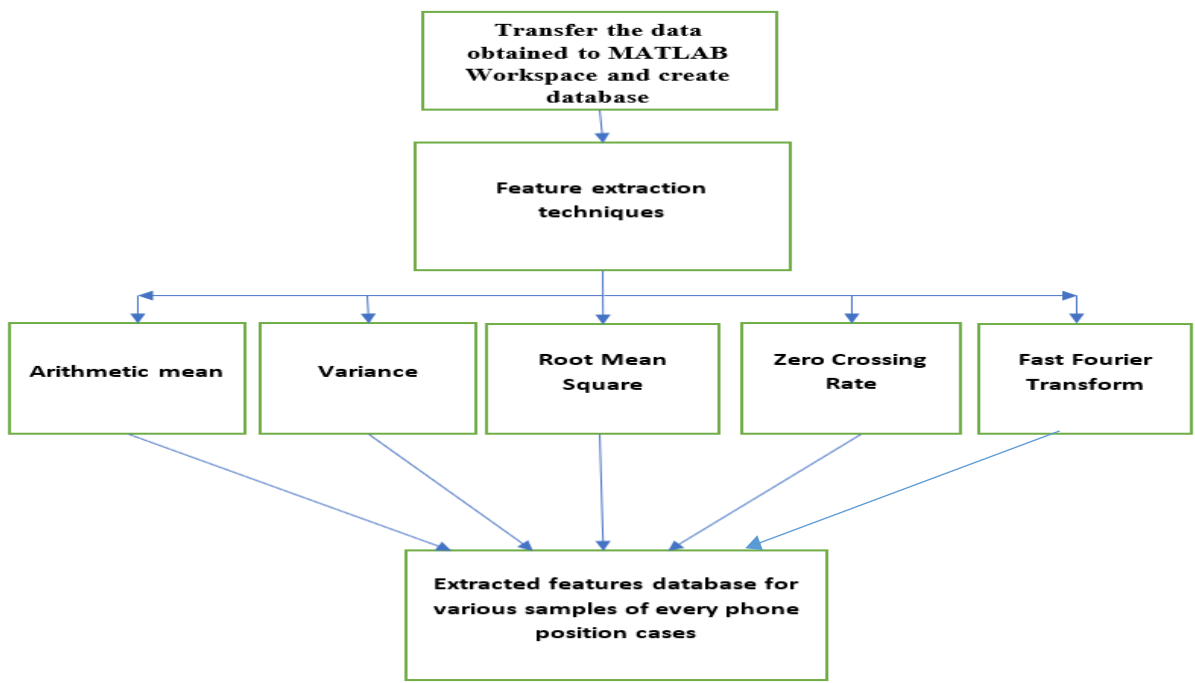

Fig.15 Feature Extraction process [17]

- Mean- the average value of magnitude of the various acceleration values and orientation values for each sample of various cases was taken for the different 3 axes.
- Variance- the average of the squared difference of the sample values from the mean value is taken independently for each of the three axes.
- Root Mean Square (RMS) – the square root of the sum of different sample values was taken independently for each of the three axes.
- Zero crossing rate (ZCR) – the number of times acceleration samples change from positive to negative and back is calculated.
- Fast Fourier transform (FFT) - the first 5 of the fast- Fourier transform coefficients of the various acceleration samples for the three axes are taken. Only five samples are chosen because they capture the main frequency components and the use of additional coefficients does not improve the accuracies.

From the database created, the number of acceleration related features extracted for a sample of a case for only one axis is 9. And when all the three axes are considered the total number of acceleration related features extracted for each sample is 27.Similarly considering all the axes and total number of orientation related features extracted for each of the sample of a particular case is 27.Combining the acceleration related features extracted along with the orientation related features, the total number no of features obtained for each sample of a particular case is 54.

### 4. CREATION OF EXTRACTED FEATURE DATABASE

Features were extracted based on the sample values for different phone slip cases for the phone acceleration in the three axes and the phone orientation along the three axes. For each sample for a particular case, a total of 54 features were extracted. The same procedure was repeated for the other samples of different cases also. To use these extracted features for further classification techniques, a database was created consisting of all the features segregated in an appropriate form.

The features were arranged based on the samples values for different phone slip cases [18]. The features were also arranged in a manner where all the features for all samples for one particular case was put together. The reason to create this database to perform preliminary checks to check if the features extracted is valid or not. To check the validity of the feature database created, the concept of correlation was used. It was checked if the correlated value tended to 1 when different samples of same phone slip case was considered. It was also checked if the correlated value tended to 0 when different sample value from different case was considered. This was done to check that all the sample value for each case were similar to each other or not and to check if the samples value for different cases were not similar to each other. Once the various samples were checked, it was then subjected to various classification techniques.

### 5. RESULT

In our paper, six phone slip cases were considered, namely normal touch keep (A), accidental keep (B), complete slip (C), slip till tipping point (D), flip (E) and fall (F). The acceleration and orientation related features for various samples for each case was taken. Various feature extraction techniques were employed on these samples to obtain the required database. On the values present in the database, various classification algorithms were used to classify or distinguish these cases. The number of training and testing feature has been is presented in table 6.1

**Table 1.Feature split up table**

| Total Case | 6 |
|---|---|
| Total Features | 6480 |
| Training Features | 4536 |
| Testing Features | 1944 |
| Samples/case | 20 |

We uses four major neural network classifier to observer the results they are Feedforward[19], Pattern NET, Fit NET, and Cascade NET Pattern Recognition network uses basic neural network for grouping or classifying the patterns present in the data set. In this paper we used pattern recognition for classifying the pattern among different slip case and training has been done by proving predefined targets in supervised way .Fitness network is basic feedforward network which uses genetic algorithm to tune the learning parameter ,this network can be used for solving both regression as well as for classification .Here it is used for classification purpose by providing 70% data for training and the rest 30% for testing .In Cascaded neural Network each layer neurons are inter-connected with its preceding ,succeeding, and input layer neurons which produces confined outputs .since the network has more number of weights to be updated due to more interconnection neurons ,the memory and time required to training is comparatively higher than other networks

**Table 2. Classification Performance**

| Cases | Pattern Net (%) | Feedforward (%) | Fit Net (%) | Cascade (%) | Average on cases (%) |
|---|---|---|---|---|---|
| AB | 66.67 | 58.33 | 41.66 | 58.33 | 56.247 |
| AC | 66.67 | 66.66 | 50 | 58.33 | 60.415 |
| AD | 91.66 | 50 | 50 | 75 | 66.665 |
| AE | 83.33 | 50 | 50 | 58.33 | 60.415 |
| AF | 100 | 50 | 58.33 | 50 | 64.582 |
| BC | 66.66 | 50 | 41.66 | 41.66 | 49.995 |
| BD | 75 | 58.33 | 41.66 | 58.33 | 58.330 |
| BE | 58.33 | 58.33 | 50 | 66.66 | 58.330 |
| BF | 58.33 | 50 | 50 | 58.33 | 54.165 |
| CD | 75 | 41.66 | 50 | 41.66 | 52.080 |
| CE | 83.33 | 41.66 | 50 | 50 | 56.247 |
| CF | 58.33 | 50 | 50 | 41.66 | 49.997 |
| DE | 50 | 50 | 41.66 | 58.33 | 49.997 |
| DF | 58.33 | 41.66 | 41.66 | 66.66 | 52.077 |
| EF | 58.33 | 58.33 | 50 | 50 | 54.165 |
| Average | 69.998 | 51.664 | 47.775 | 55.552 | 56.247 |

The total phone slip cases considered is six and the total samples taken per case is 20. The total features extracted for all the samples in every case is 6480. The total testing and training features are 1944 and 4536 respectively as shown in Table.1.

On the total features extracted, various classification algorithms as mentioned above were implemented. The results obtained after implementing these neural networks on the extracted features are summarized in classification performance table (table 2). In table 2, various neural network was implemented on pairs of the phone slip cases and the classification accuracy in percentage were observed. Then, for the particular neural network, the average of classification percentages was calculated to indicate which of the neural network can classify various phone slip cases. Also, the average of the classification percentage for each sample after implementing in the various neural networks were calculated to identify which phone slip case can be easily classified. From table 2, it can be inferred that the classification accuracies in percentages among various pairs of the phone slip cases is maximum when pattern net neural network is employed because the classification percentages for a particular pair after employing every neural network is maximum in Pattern Net and the average of the classification percentages for various cases is maximum in Pattern Net. Therefore, the minimum classification accuracy is obtained in Fit Net followed by Feed- forward, cascade and the maximum classification accuracy is obtained in Pattern Net neural network. Moreover the recognition on AD shows more on an average when compare to other cases due to high accuracy rate in Pattern Net .The highest 100% accuracy can be found in Pattern Net on AF samples since cascaded Net show 25% lesser then AD the average on cases reduces by 2.083 units .In some case like DF, BE and DE where Cascade Net perform better the other networks of accuracy 66.66% ,66.66% and 58.33% respectively .Finally the overall ranking of four network has been given as 1-Pattern Net, 2-Cascade Net, 3-Feedforward Net and 4-FitNet. Since Fit Net has learning factor in both neural net and GA therefore the accuracy of Fit net can be increased by increasing the training samples.

### 6 CONCLUSION

From the results obtained, we can conclude that after employing the Pattern Net on the extracted features, the classification accuracies for different pairs of the phone slip case is maximum and also on the whole the average classification accuracies for pattern Net for all the cases is maximum (69.998 %). Therefore, it can be concluded that out of the four used neural networks, the Pattern net can more efficiently classify various phone slip cases so say if a particular phone slip position is vulnerable or safe.

In future, we plan to improve our phone slip recognition system in several ways. Firstly, the efficiency of the project can be improved if it can classify various complex phone positions. Secondly various additional and more sophisticated features can be extracted from various samples of different chosen cases to improve the classification accuracy. The work presented in this paper is a part of a larger effort to classify various phone positions into vulnerable and safe positions. Mobile phones are becoming increasingly advanced and the in- built sensors present in them can be configured in a way to identify if the phone is in vulnerable position to fall or it's in a safe position.

**REFERENCES**

[1] Inooka, H., Ohtaki, Y. Hayasaka, H. Suzuki, A., and Nagatomi, R. 2006. Development of advanced portable device for daily physical assessment. In SICE-ICASE, International Joint Conference, 5878-5881.
[2] Ming Liu. A Study of Mobile Sensing Using Smartphones. In International Journal of Distributed Sensor Networks, 2013.
[3] Jennifer R. Kwapisz, Gary M. Weiss, Samuel A. Moore. Activity Recognition using cell phone accelerometers: SIGKDD Explorations Volume 12, Issue 2, 74-82, 2010.
[4] Brezmes, T., Gorricho, J.L., and Cotrina, J. Activity Recognition from accelerometer data on mobile phones. In IWANN '09: Proceedings of the 10th International Work Conference on Artificial Neural Networks, 796-799, 2009.
[5] Maurer, U., Smailagic , D., & Deisher, M. Activity recognition and monitoring using multiple sensors on different body positions. In IEEE Proceedings on the International Workshop on Wearable and Implantable Sensor Networks, 3(5), 2006.
[6] Ozlem Durmaz Incel. Analysis of Movement, Orientation and Rotation-Based Sensing for Phone Placement Recognition: Open Access Sensors 2015, 15, 25474-25506, 2015.
[7] Gyorbiro, N., 2008. An activity recognition system for mobile phones. In Mobile Networks and Applications, 14(1), 82-91.
[8] Morillo, L.M.S.; Gonzalez-Abril, L.; Ramirez, J.A.O.; de la Concepcion, M.A. Low energy physical activity recognition system on smartphones. Sensors 2015, 15, 5163–5196, 2015.
[9] Coskun, D.; Incel, O.; Ozgovde, A. Phone position/placement detection using accelerometer: Impact on activity recognition. In Proceedings of the 2015 IEEE Tenth International Conference on ISSNIP, 7–9 April 2015; pp. 1–6.
[10] Bayat, K.; Pomplun, M.; Tran, D.A. A Study on Human Activity Recognition Using Accelerometer Data from Smartphones. Proced. Comput. Sci. 2014, 34, 450–457.
[11] Yunus Emre ¨Ustev. User, Device, Orientation and Position Independent Human Activity Recognition On Smart Phones, B.S., Computer Engineering, ˙Istanbul Technical University, 2008.
[12] Benish Fida, Ivan Bernabucci, Daniele Bibbo, Silvia Conforto and Maurizio Schmid. Pre-Processing Effect on the Accuracy of Event-Based Activity Segmentation and Classification through Inertial Sensors: 15, 23095-23109, 2015.
[13] Choudhury, T., Consolvo, S., Harrison, B., LaMarca, A., LeGrand, L., Rahimi, A., Rea, A., Borriello, G., Hemingway, B., Klasnja, P., Koscher, K., Landay, J., Lester, J., Wyatt, D., and Haehnel, D. 2008. The mobile sensing platform: An embedded activity recognition system. In IEEE Pervasive Computing, 7(2), 32-41.
[14] Ichikawa, F.; Chipchase, J.; Grignani, R. Where's the Phone? A Study of Mobile Phone Location in Public Spaces. In Proceedings of the 2005 2nd International Conference on Mobile Technology, Applications and Systems, , 2005
[15] Inooka, H., Ohtaki, Y. Hayasaka, H. Suzuki, A., and Nagatomi, R. 2006. Development of advanced portable device for daily physical assessment. In SICE-ICASE, International Joint Conference, 5878-5881.
[16] https://in.mathworks.com/products/matlabmobile.html#acquire-data-from-sensors
[17] Lester, J., Choudhury, T. and Borriello, G. 2006. A practical approach to recognizing physical activities. Lecture Notes in Computer Science: Pervasive Computing, 1–16.
[18] Krishnan, N., Colbry, D., Juillard, C., and Panchanathan, S. 2008. Real time human activity recognition using tri-Axial accelerometers. In Sensors, Signals and Information Processing Workshop.
[19] Eric A. Plummer, Time Series Forecasting With Feed-forward Neural Networks: Guidelines and Limitations, July 2000